\documentclass[sigconf]{acmart}

\AtBeginDocument{%
  \providecommand\BibTeX{{%
    \normalfont B\kern-0.5em{\scshape i\kern-0.25em b}\kern-0.8em\TeX}}}

\renewcommand\footnotetextcopyrightpermission[1]{}

\acmConference[Workshop on Workplace Wellbeing, CHI'24]{ACM Conference}{May 11, 2024}{Honolulu, HI, USA}%

\begin{document}

\title{The Adaptive Workplace: Orchestrating Architectural Services around the Wellbeing of Individual Occupants}
\renewcommand{\shorttitle}{The Adaptive Workplace}


\author{Andrew Vande Moere}
\email{andrew.vandemoere@kuleuven.be}
\orcid{0000-0002-0085-4941}
\affiliation{
  \institution{Research[x]Design, \break Dept. of Architecture, KU Leuven}
  \streetaddress{Kasteelpark Arenberg 1 - box 2431}
  \city{Leuven}
  \country{Belgium}
  \postcode{3001}
}

\author{Sara Arko}
\email{sara.arko@iri.uni-lj.si}
\affiliation{
  \institution{Institute for Innovation and Development of University of Ljubljana (IRI UL)}
  \city{Ljubljana}
  \country{Slovenia}
}

\author{Alena Safrova Drasilova}
\email{alena.drasilova@mail.muni.cz}
\affiliation{
  \institution{Department of Business Management, Masaryk University}
  \city{Brno}
  \country{Czech Republic}
}

\author{Tomáš Ondráček}
\email{ondracek.t@mail.muni.cz}
\affiliation{
  \institution{Department of Business Management, Masaryk University}
  \city{Brno}
  \country{Czech Republic}
}

\author{Ilaria Pigliautile}
\email{ilaria.pigliautile@unipg.it}
\affiliation{
  \institution{Environmental Applied Physic Lab, Engineering Department, University of Perugia}
  \city{Perugia}
  \country{Italy}
}

\author{Benedetta Pioppi}
\email{rad@valerielettrica.com}
\affiliation{
  \institution{EValTech, R\&D Department of Elettrica Valeri srl}
  \city{Perugia}
  \country{Italy}
}

\author{Anna Laura Pisello}
\email{anna.pisello@unipg.it}
\affiliation{
  \institution{Environmental Applied Physic Lab, Engineering Department, University of Perugia}
  \city{Perugia}
  \country{Italy}
}

\author{Jakub Prochazka}
\email{jak.prochazka@mail.muni.cz}
\affiliation{
  \institution{Department of Business Management, Masaryk University}
  \city{Brno}
  \country{Czech Republic}
}

\author{Paula Acuna Roncancio}
\email{paula.acuna.roncancio@deltalight.com}
\affiliation{
  \institution{Delta Light NV}
  \city{Wevelgem}
  \country{Belgium}
}

\author{Davide Schaumann}
\email{d.schaumann@technion.ac.il}
\affiliation{
  \institution{Intelligent Place Lab, Faculty of Architecture and Town Planning, Technion – Israel Institute of Technology}
  \city{Haifa}
  \country{Israel}
}

\author{Marcel Schweiker}
\email{mschweiker@ukaachen.de}
\affiliation{
  \institution{Institute for Occupational, Social and Environmental Medicine, Medical Faculty, RWTH Aachen University}
  \city{Aachen}
  \country{Germany}
}

\author{Binh Vinh Duc Nguyen}
\email{alex.nguyen@kuleuven.be}
\orcid{0000-0001-5026-474X}
\affiliation{
  \institution{Research[x]Design, \break Dept. of Architecture, KU Leuven}
  \streetaddress{Kasteelpark Arenberg 1 - box 2431}
  \city{Leuven}
  \country{Belgium}
  \postcode{3001}
}

\renewcommand{\shortauthors}{Vande Moere et al.}

\begin{abstract}
    As the academic consortia members of the EU Horizon project SONATA ("Situation-aware OrchestratioN of AdapTive Architecture"), we respond to the workshop call for "\textit{Office Wellbeing by Design: Don’t Stand for Anything Less}" by 
    proposing the "Adaptive Workplace" concept. 
    In essence, our vision aims to adapt a workplace to the ever-changing needs of individual occupants, instead of that occupants are expected to adapt to their workplace.
\end{abstract}

\keywords{human-building interaction (HBI), smart office, adaptive architecture, robotic furniture, human-robot interaction (HRI)}

\begin{teaserfigure}
  \includegraphics[width=\textwidth]{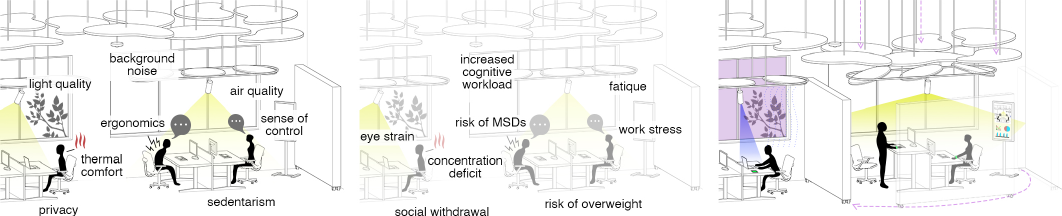}
  \caption{
  Our vision of the "Adaptive Workplace" proposes that: 1) potential or actual issues that might negatively affect the wellbeing of occupants are identified in a \textbf{contextual} way. These issues are then: 2) addressed by \textbf{multiple}  architectural services on different building layers; that 3) are jointly \textbf{orchestrated} as one holistic action; which 4) is \textbf{equitably} distributed over the individual occupants according to their personal preferences. 
  }
  \Description{xxx.}
  \label{fig:teaser}
\end{teaserfigure}

\maketitle

\section{The Problem}
As the academic consortium members of the EU Horizon RIA project SONATA (short for "Situation-aware OrchestratioN of AdapTive Architecture"), this short positional paper reacts to the workshop call "\textit{Office Wellbeing by Design: Don’t Stand for Anything Less}". 
We answer the workshop's call to "\textit{formulate a research agenda to target the grand challenges in workplace wellbeing and develop actionable measures to translate research into practice}" by proposing the grand vision of the "Adaptive Workplace". 

\subsection{Health \& Wellbeing in the Open-Plan Office}
Most knowledge-based industries have adopted the "open-plan" or "activity-based" workplace layout to host the wide variety of concurrent work activities that arise because their workforce is only intermittently and unpredictably present. 
Post-occupancy evaluations yet show that the majority of workers feel dissatisfied with at least one aspect of their workplace design, which is twice as problematic for open-plan than for enclosed office layouts. Because their complaints typically focus on acoustics (e.g. people talking, speech privacy, phone calls), a perceived lack of control, insufficient space and privacy-related concerns \cite{Parkinson2023}, it is believed that their health, wellbeing \cite{Rashid2008}, productivity and social relations are harmed, in terms of individuals as well as the whole overarching work organisation \cite{Hodzic2021,Borge2023}.


\subsection{Architecturally Adaptive Technology}

\begin{figure}[ht]
  \includegraphics[width=0.7\linewidth]{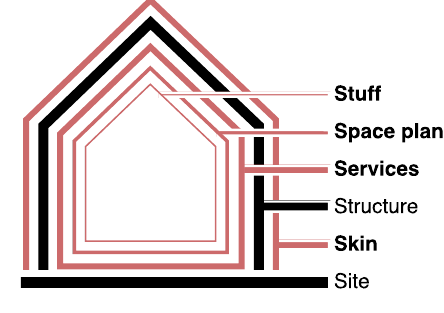}
  \caption{The six shearing layers of a building on which adaptive architectural technology could be applied.
  }
  \label{fig:layers}
\end{figure}

Due to its inherent shared nature, an ideal open-plan workplace design that works for everyone and all the time might not exist. Instead, a workplace design that needs to cope with multiple work conditions must allow occupants to cope with negative work-related situations in a resilient way \cite{Forooraghi2021}. So-called `adaptive' workplace technologies try to address this goal by interpreting automated and voluntary data streams to control a wide range of architectural services. Divided by the different `shearing layers' of a building \cite{Brand1994} (see \autoref{fig:layers}), prototypical examples of such `adaptive' architectural services that are already commercially available include, but are not limited to: `automatic' shading systems that protect the interior from outdoor environmental conditions (skin); `personalised' Heating, Ventilation, Air conditioning and Cooling (HVAC) and `smart' lighting systems that optimise human thermal, visual, and respiratory comfort conditions (services); `flexible' office partitions that can be manually arranged to spatially demarcate different work conditions (space plan); and `height-adjustable' desks that help reduce sedentary behaviour (stuff). 

However, post-occupancy studies keep demonstrating how adaptive architectural services cannot ameliorate all negative health concerns, mainly because they fail to adequately address the complex interplay of negative risk factors (e.g. density, acoustic, visual disturbances) \cite{Nag2019}, operate agnostically of individual preferences \cite{Luo2018}, and fail to exploit health benefits that could lead to higher human adaptive capacity and resilience \cite{Schweiker2022}. 
The manual operation of adaptive architectural services is often underused because it is considered exhausting, time-consuming or socially embarrassing \cite{Zamani2019}, while communally shared building control systems tend to be experienced as being impractical or socially awkward \cite{Schweiker2016}. 
Fully autonomous building control systems are, excluding a few exceptions that are only found in academic research, solely designed to optimise towards averaged human comfort and environmental satisfaction criteria (\autoref{fig:teaser}A), completely overlooking the dynamic, contextual and individual nature of human comfort requirements that change among the work activities during a single workday, and even significantly differ between the multiple occupants that share the same workplace \cite{Parkinson2021}. 
Consequently, even when most shared workplaces are already equipped with various adaptive architectural technologies like sunshades, HVAC outlets or lighting fixtures, individual workers lack the knowledge on how to operate them effectively in shared settings, while work organisations and building certification institutions do not possess persuasive evidence on how to recommend or prescribe them.

\section{The Solution}
Knowing that the physical design of a shared workplace influences the health and wellbeing of its occupants \cite{Roskams2020}, our general premise is that adaptive architectural services have now become sufficiently mature to holistically and constantly adapt a workplace to the ever-changing needs of its occupants, instead of that occupants need to adapt to their workplace.
\begin{itemize}
    \item \textbf{Multi-layered.} Realising that a single adaptive architectural service cannot rectify all common dissatisfactions, we propose that architectural adaptations should be deployed on multiple building shearing layers simultaneously. 
    \item \textbf{Situational-aware.} To be able to address the interplay of concurrent work conditions within the same shared workplace, we propose that architectural adaptation should not react to averaged human comfort requirements but rather to `situations', defined as a dynamic combination of environmental conditions (e.g. noise, glare), work activities (e.g. focused, social interaction) and social preferences, from individuals and vulnerable groups that can be characterised by gender, demographics or previous experience with the workplace.
    \item \textbf{Orchestrated.} We then foresee that each adaptive layer should leverage the physical manifestation of all other layers in both time and space, insofar that their holistic effect on health and wellbeing should be more beneficial than each individual layer separately.
    \item \textbf{Equitable.} Furthermore, each adaptation should negotiate its estimated impact between the multiple situations that coexist in the same shared workplace, insofar that it could prioritise situations that are in highest need.
\end{itemize}


\section{Future Research Agenda}
In the context of human-computer interaction domain, the proposed "Adaptive Workplace" vision requires a research agenda that is spread over several domains. 

\subsection{Measuring Impact}
We propose that the impact of individual as well as combination of adaptive technologies on occupants should become objectively benchmarked. This endeavour should reach well beyond simply capturing environmental measures (e.g. ambient temperature, humidity, air quality) and instead focus on how occupants perceive, feel and interpret the actual effect of the adaptations, knowing that individual preferences can change over time, and that individual preferences cannot always be met in a 'shared' setting. Such a benchmarked should also include an socio-economic short- to longer-term evaluation, including a cost-benefit analysis.

\subsection{Designing Novel Adaptive Technologies}
We believe there still exists a largely unexplored design space of adaptive architectural services that could be invented, prototyped and tested. This design space includes the conceptualisation of novel physical embodiments as well as how occupants can be empowered to take control over them. Despite significant insights in human-robotic interaction in general, and robotic furniture more specifically, commonly existing adaptive architectural services are seemingly not progressing beyond fully integrated, and therefore inflexible, building systems of which the effect spans a whole room. Instead, novel adaptive technologies should become more flexible and modular, while their spatial resolution should target an individual occupant. 

\subsection{Controlling Adaptive Architectural Services}
Current adaptive architectural services tend to be controlled by relatively archaic wall-mounted dials or screens or relatively impracticable smartphone applications that do not invite occupants to take control nor allow any "smart" system to learn from how occupants actually perceive its performance.
Instead, we think that controlling an adaptive workplace should be situational in nature, so that it does not interrupt the natural workflow of occupants. To ensure that its "autonomous" actions are trusted and accepted, new "human-in-the-loop" interfaces need to be developed that allow occupants to be transparently informed about the decisions that underlie its operation.

\subsection{Orchestrating the Impact of Adaptive Architectural Services}
The adaptive workplace requires the real-time orchestration of different adaptive architectural services that considers the non-linear relation between spatial configurations and social outcomes. Simulation and optimisation frameworks are needed to predict, analyse, and compare the outcomes of alternative orchestration strategies on the health and well-being of individual occupants that are copresent in the same space. 
These simulations should combine architectural knowledge with artificially intelligent algorithms capable of identifying optimal solutions to a given situation. Furthermore, the optimisation algorithm should learn from the feedback from individual occupants to situations.

\section{Conclusion}
This short positional paper introduced the current problem of open-plan shared offices, in terms of how current architectural adaptive technologies in general fail to address the health and wellbeing issues of individual occupants. It proposed a "grand vision" of the "adaptive workplace", which is able to adapt around the needs of individual occupants instead of them having to adapt to their office. We also proposed a succinct research agenda that could underpin this grand vision.

\begin{acks}


The research presented in this paper was carried out as part of the HORIZON-HLTH-2023-ENVHLTH-02 SONATA project. This project has received funding from the European Union's Horizon Europe Research and Innovation Actions programme under Grant Agreement no. 101137507. Views and opinions expressed are however those of the authors only and do not necessarily reflect those of the European Union or Health and Digital Executive Agency. Neither the European Union nor the granting authority can be held responsible for them.


\end{acks}

\bibliographystyle{ACM-Reference-Format}
\bibliography{sample-base}


\end{document}